%% file: main.tex
\documentclass[sigconf,natbib=false]{acmart}

\usepackage{amssymb}
\usepackage{multirow}
\usepackage{multicol}
\usepackage{listings}
\usepackage{adjustbox}
\usepackage{makecell}
\usepackage{enumitem}
\usepackage{colortbl}
\usepackage{booktabs}
\definecolor{ForestGreen}{RGB}{34,139,34}
\usepackage{algorithm}
\usepackage[noend]{algpseudocode}
\usepackage{tabularx}
\usepackage{pifont}
\usepackage{graphicx}
\usepackage{duckuments}
\usepackage{float}
\usepackage{cancel}
\usepackage[frozencache,cachedir=minted-cache]{minted} 

\AtBeginDocument{%
  \providecommand\BibTeX{{%
    \normalfont B\kern-0.5em{\scshape i\kern-0.25em b}\kern-0.8em\TeX}}}

\copyrightyear{2024}
\acmYear{2024}
\setcopyright{rightsretained}
\acmConference[FPGA '24]{Proceedings of the 2024 ACM/SIGDA International Symposium on Field Programmable Gate Arrays}{March 3--5, 2024}{Monterey, CA, USA}
\acmBooktitle{Proceedings of the 2024 ACM/SIGDA International Symposium on Field Programmable Gate Arrays (FPGA '24), March 3--5, 2024, Monterey, CA, USA}
\acmDOI{10.1145/3626202.3637569}
\acmISBN{979-8-4007-0418-5/24/03}

\newcounter{BalanceAtReference}
\newcounter{ReferenceIndexForBalancing}

\makeatletter

\global\@ACM@balancefalse

\def\@balancelastpageonce{%
  \ifnum\value{ReferenceIndexForBalancing}=\value{BalanceAtReference}
    \newpage
  \else
    \relax
  \fi
  \stepcounter{ReferenceIndexForBalancing}
}
\pretocmd{\bibitem}{\@balancelastpageonce}
  {} 
  {\@latex@error{Patching \bibitem failed}{\@ehd}}

\makeatother

\makeatletter
\gdef\@copyrightpermission{
   Permission to make digital or hard copies of part or all of this work for personal or classroom use is granted without fee provided that copies are not made or distributed for profit or commercial advantage and that copies bear this notice and the full citation on the first page. Copyrights for third-party components of this work must be honored. For all other uses, contact the owner/author(s).
}
\makeatother

\newcommand{\M}{SSR}
\newcommand{\jz}{\textcolor{black}}
\usepackage{tikz}

\begin{document}

\title{~\M: Spatial Sequential Hybrid Architecture for  Latency Throughput Tradeoff in Transformer Acceleration
}
\pagenumbering{arabic}

\settopmatter{authorsperrow=4}
\author{Jinming Zhuang}
\affiliation{%
  \institution{\small{University of Pittsburgh} 
  \country{USA}}}
\email{jinming.zhuang@pitt.edu}

\author{Zhuoping Yang}
\affiliation{%
  \institution{\small{University of Pittsburgh} \country{USA}}}
\email{zhuoping.yang@pitt.edu}

\author{Shixin Ji}
\affiliation{%
  \institution{\small{University of Pittsburgh} \country{USA}}}
\email{shixin.ji@pitt.edu}

\author{Heng Huang}
\affiliation{%
  \institution{\small{University of Maryland} \country{USA}}}
\email{heng@umd.edu}

\author{Alex K. Jones}
\affiliation{%
  \institution{\small{University of Pittsburgh} \country{USA}}}
\email{akjones@pitt.edu}

\author{Jingtong Hu}
\affiliation{%
  \institution{\small{University of Pittsburgh} \country{USA}}}
\email{jthu@pitt.edu}

\author{Yiyu Shi}
\affiliation{%
  \institution{\small{University of Notre Dame} \country{USA}}}
\email{yshi4@nd.edu}

\author{Peipei Zhou}
\affiliation{%
  \institution{\small{University of Pittsburgh}
  \country{USA}}}
\email{peipei.zhou@pitt.edu}

\renewcommand{\shortauthors}{Jinming Zhuang et al.}

\input{0_abstract.tex}

\ccsdesc[500]{Computer systems organization~Heterogeneous (hybrid) systems}
\ccsdesc[500]{Hardware~Hardware-software codesign}

\keywords{Heterogeneous Computing, Domain-Specific Accelerator, Versal ACAP,  Transformers, Design Space Exploration, Latency Throughput Tradeoff, Deep Learning}


\maketitle



\input{1_Introduction}
\input{2_DesignChallenge}

\input{3_RelatedWork}
\input{4_SSR_Methodology}
\input{6_SSR_Experiments}
\input{7_Discussion}
\input{8_Conclusion}


\bibliographystyle{unsrt}
\bibliography{reference}

\end{document}

%% file: 0_Abstract.tex
\begin{abstract}
\label{sec:abstract}
With the increase in the computation intensity of the chip, the mismatch between computation layer shapes and the available computation resource significantly limits the utilization of the chip. 
Driven by this observation, prior works discuss spatial accelerators or dataflow architecture to maximize the throughput. 
However, using spatial accelerators could potentially increase the execution latency. 
In this work, we first systematically investigate two execution models: (1) sequentially (temporally) launch one monolithic accelerator, and (2) spatially launch multiple accelerators. 
From the observations, we find that there is a latency throughput tradeoff between these two execution models, and combining these two strategies together can give us a more efficient latency throughput Pareto front. 
To achieve this, we propose spatial sequential architecture (\M) and ~\M~ design automation framework to explore both strategies together when deploying deep learning inference. 
We use the 7nm AMD Versal ACAP VCK190 board to implement SSR accelerators for four end-to-end transformer-based deep learning models. 
~\M~ achieves average throughput gains of 2.53x, 35.71x, and 14.20x under different batch sizes compared to the 8nm Nvidia GPU A10G, 16nm AMD FPGAs ZCU102, and U250.
The average energy efficiency gains are 8.51x, 6.75x, and 21.22x, respectively.
Compared with the sequential-only solution and spatial-only solution on VCK190, our spatial-sequential-hybrid solutions achieve higher throughput under the same latency requirement and lower latency under the same throughput requirement.
We also use ~\M~ analytical models to demonstrate how to use ~\M~ to optimize solutions on other computing platforms, e.g., 14nm Intel Stratix 10 NX.
\end{abstract}

%% file: 1_Introduction.tex
\section{Introduction}
\label{sec:introduction}
\label{sec:challenges}
Latency and throughput are two crucial performance metrics when deploying deep learning models on various computing platforms. 
Depending on the nature of the applications and different user expectations, different application scenarios have different latency requirements. 
For example, the latency requirement in autonomous driving is more stringent than that in video conferencing. 
The former requires milliseconds or submillisecond latency~\cite{web-self-driving,cernNews,zhang2019accelerating,wp505_versal_acap} for a life-critical system whereas the latter has a looser latency requirement of hundreds of milliseconds.
Furthermore, throughput is also needed to be considered. 
For example, in data center services, e.g., Microsoft~\cite{putnam2014reconfigurable,caulfield2016cloud,firestone2018azure,fowers2018configurable}, Google~\cite{jouppi2017datacenter}, AWS~\cite{awsinferentia}, etc, higher throughput means less amount of data center servers and therefore less power consumption for the same workload. 
On the other hand, it can also support more volumes of users while ensuring real-time user content updates with the same amount of servers. 
For autonomous vehicles, to safely navigate the changing environments, higher throughput means processing higher amounts of sensor data to make real-time decisions~\cite{lu2023vehicle}. 

The two factors are also intertwined and there is a design tradeoff between latency and throughput. 
In general cases, a system can not get high throughput and low latency simultaneously. 
If a design requires higher throughput which can be achieved by batching more data, the system would have to sacrifice latency.
While users can only explore latency throughput tradeoff by changing the batch size when using the off-the-shelf deep learning framework on GPUs, FPGA accelerators~\cite{zhou2023refresh,zhang2018dnnbuilder,eftrainTODAES22,fpga21tecs,MPOPU} and other tiled accelerators~\cite{MaKaishengISCA23,kim2023full,charmFPGA23,aimICCAD2023,yang2023challenges,yan2022computing,yan2022swim,yan2023improving} provide more flexibility and users have a larger design space to explore the latency throughput tradeoff.

\setlength{\textfloatsep}{2pt}
\begin{figure}[tb]
\centering
\includegraphics[width=1\columnwidth]{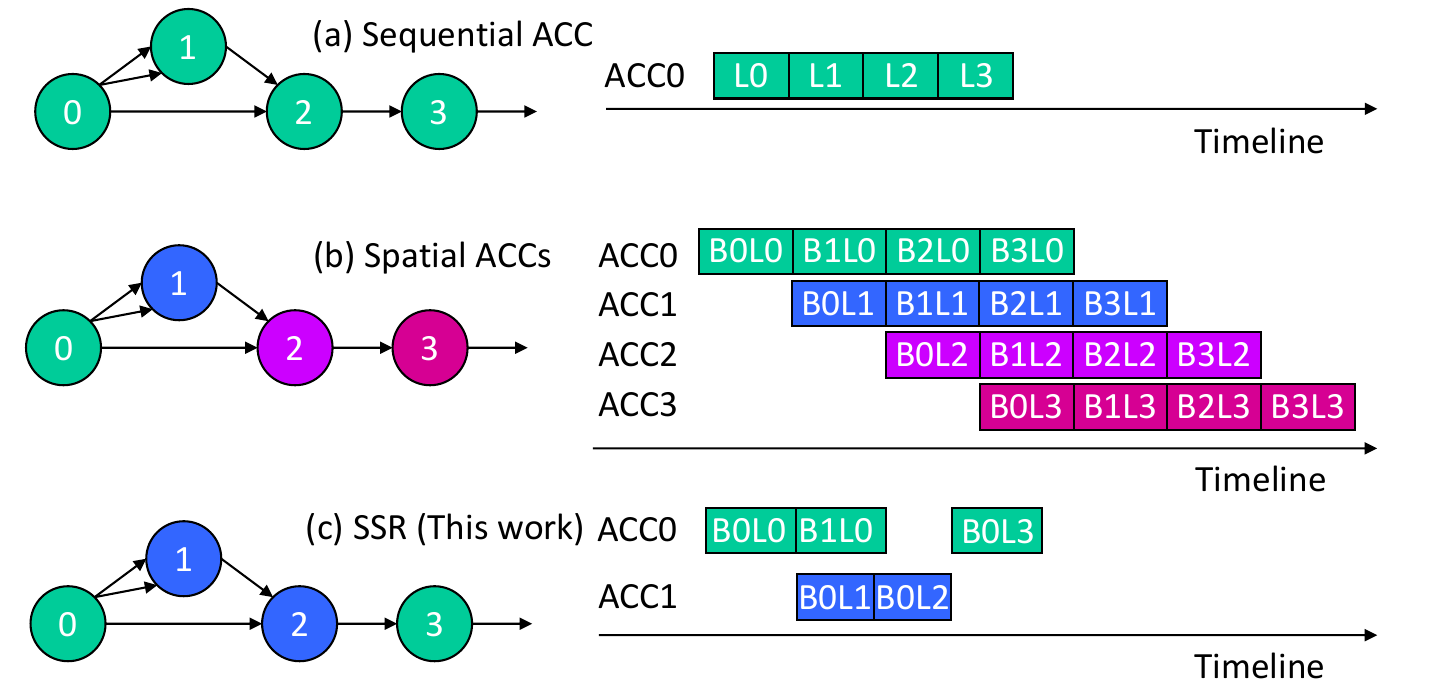}
\caption{Execution models for sequential, spatial, and our proposed spatial-sequential-hybrid architecture (SSR).}
\label{fig:exec_model}
\vspace{-10pt}
\end{figure}

\begin{figure}[tb]
\centering
\includegraphics[width=1\columnwidth]{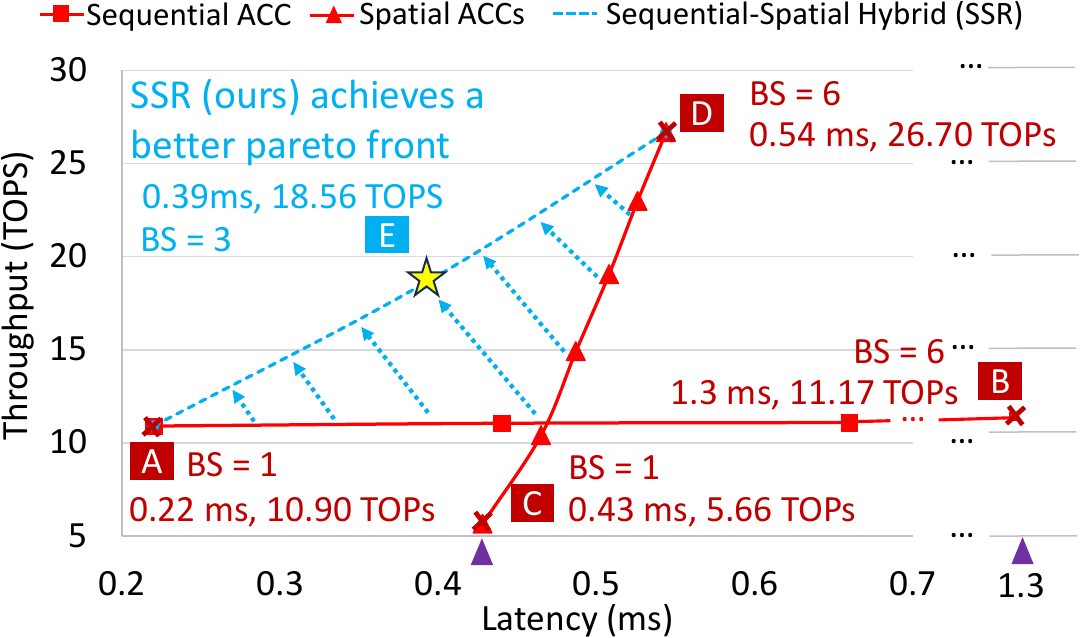}
\vspace{-15pt}
\caption{Latency and throughput tradeoff under different strategies for one representative vision transformer model, i.e., DeiT-T~\cite{deit}. ~\M~ (ours) achieves a better Pareto front than sequential acc and fully spatial accs designs.}
\label{fig:tradeoff}
\end{figure}

By using on-chip local scratchpad memory and configurable processing elements, users can design customized accelerators (accs) that fit certain computations, and this is called accelerator (acc) customization.
There are different strategies when mapping multiple layers within a deep learning model onto FPGAs or tiled accelerators. 
One common method is to design one unified acc that can compute different layers within the model graph and the unified accelerator is launched \textbf{sequentially} to finish all the layers~\cite{zhang2018caffeine}.
The execution model timeline is shown in Figure~\ref{fig:exec_model}(a). 
Here we use a graph with four layers \texttt{0}-\texttt{3} to illustrate.  
The arrows show the layer dependencies in the graph.
Where there is only one acc, \texttt{ACC0}, four layers \texttt{L0}, \texttt{L1}, \texttt{L2}, \texttt{L3} are launched sequentially while honoring the dependencies in the graph.
When users increase the batch size, in the timeline, \texttt{L0-L3} will become longer. 
We also apply the sequential acc design strategy and map one representative deep learning application, an INT8 quantized vision transformer model DeiT-T~\cite{deit} for image classification task on AMD ACAP VCK190~\cite{versal_vck190}. 
We sweep the batch size from 1 to 6 and find the customized monolithic acc that gives the highest throughput under each batch size. 
We plot the latency and corresponding throughput for each batch size as a 2-D scatter plot and add the trendline as shown in Figure~\ref{fig:tradeoff}. 
From point \texttt{A} to point \texttt{B}, the latency increases from 0.22 ms to 1.3 ms. 
The effective throughput slightly increases from 10.90 TOPS to 11.17 TOPS, which means the sequential acc design strategy achieves 10.9\% utilization of the peak INT8 computation performance (102 TOPS) for AMD VCK190.
The underlying reasons for such a utilization are: (1) the computation and communication patterns for different layers in DeiT-T vary a lot; (2) there is a huge mismatch between the small matrix multiply layer shape and the huge computation resource.
Therefore, the first question arises: \textit{Can we achieve a higher throughput?}

A common solution is to apply an alternative design strategy, i.e., implementing \textbf{spatial accs}~\cite{fowers2018configurable} and mapping each layer with a dedicated specialized acc, i.e., fully spatial acc design. 
The corresponding execution model timeline is shown in Figure~\ref{fig:exec_model}(b), where there are four accs,~\texttt{ACC0}-\texttt{ACC3}.
Since there are dependencies between layers 0-3, four layers in the same batch data \texttt{B0L0}, \texttt{B0L1}, \texttt{B0L2}, \texttt{B0L3} have to be launched sequentially. 
As can be observed from Figure~\ref{fig:exec_model}(b), if there is only one batch, ~\texttt{ACC0}-\texttt{ACC3} will be severely underutilized.
However, when there are more batches, e.g., \texttt{B1}-\texttt{B3}, the executions for different layers from different batches can be pipelined. 
Therefore, the utilization of ~\texttt{ACC0}-\texttt{ACC3} is greatly improved. 
This also matches the trendline in Figure~\ref{fig:tradeoff} from point \texttt{C} with throughput as 5.66 TOPS to point \texttt{D} with throughput improved to 26.70 TOPS.

When choosing from these two strategies, sequential vs. spatial, the optimal design varies under different design constraints. 
For example, in Figure~\ref{fig:tradeoff}, if the latency requirement is 0.43 ms, sequential acc is more favorable than spatial acc as point \texttt{A} achieves a higher throughput and a smaller latency than point \texttt{C}.
This is intuitive to understand. When the batch size is 1, as each spatial acc has a smaller resource than the one monolithic acc, each layer takes longer execution time on separate spatial accs than on one monolithic acc.
However, if the latency requirement is 1.3 ms, spatial acc is more favorable than sequential acc as point \texttt{D} achieves a higher throughput and smaller latency than point \texttt{B}. 
This is also intuitive to understand. 
When the batch size is large, spatial accs tend to have better customization and more batches fill the pipeline gaps and improve the utilization. 
Based on this observation, one follow-up question arises: \textit{Can we combine sequential acc and spatial acc strategies together and gain the best of both worlds?} 

Our answer is ``Yes". 
The key idea is to enable more scheduling flexibility to map any layers to any accs where the number of accs can be one to the maximum number of layers.
We illustrate such a sequential-spatial hybrid architecture (SSR) in  Figure~\ref{fig:exec_model}(c).
In this approach, there are two accs, ~\texttt{ACC0} and ~\texttt{ACC1}. Layer 0 and 3 map to acc0. Layer 1 and 2 map to acc1.
By using such hybrid architecture, users can find an even better throughput than sequential acc and spatial acc strategies. 
For example, in Figure~\ref{fig:tradeoff}, if the latency requirement is 0.43 ms, the SSR hybrid strategy (point \texttt{E}) achieves throughput 18.56 TOPS, which is 1.70x throughput improvement than the sequential acc strategy (point \texttt{A}) and 3.28x than the spatial acc strategy (point \texttt{C}).  
\textbf{{The new design points enabled by the SSR strategy constitute a better Pareto front in latency throughput tradeoff.}}
That is, our SSR sequential spatial hybrid solutions achieve higher throughput under the same latency requirement or lower latency under the same throughput requirement compared with the sequential-only solution and spatial-only solution. In summary, our main contributions are:
\begin{itemize}[leftmargin=*]
\item \textbf{Design Challenges Analysis:}
To understand the performance, we first perform an in-depth kernel profiling of DeiT-T on Nvidia GPU A10G in Section~\ref{sec:challenge}. 
Then we discuss the challenges of exploring latency throughput tradeoff for deep learning applications and propose our design principles.
\item \textbf{~\M~ Accelerator and Framework:}
We propose ~\M~ accelerator, a novel sequential and spatial hybrid accelerator template, and ~\M~ framework, a programming mapping solution, in Section~\ref{sec:method} to leverage the ACAP's heterogeneous components within the same system-on-chip, including FPGA and AIE vector cores. 

\item \textbf{~\M~ Implementations:} 
{We deploy the ~\M~ framework to explore latency throughput tradeoff of four models on VCK190 in Section~\ref{sec:results}. 
Our on-board experiments demonstrate that under various latency constraints, ~\M~ achieves
average throughput gains as 2.53x, 35.71x, and 14.20x under different batch sizes compared to the 8nm Nvidia GPU A10G, 16nm AMD FPGAs ZCU102, and U250.
The average energy efficiency gains are 8.51x, 6.75x, and 21.22x, respectively. 
}
\item \textbf{Open-source Tools and Discussions on Mapping Insights}:
We open-source our tools with detailed guides to reproduce all of the results presented in this paper: \textcolor{blue}{\url{https://github.com/arc-research-lab/SSR}}.
We also discuss mapping insights in Section~\ref{sec:insight}.

\end{itemize}




%% file: 2_DesignChallenge.tex
\vspace{-5pt}
\section{Design Challenges and Proposed Solution}
\label{sec:challenge}

\begin{figure}[h]
\centering
\includegraphics[width=0.8\columnwidth]{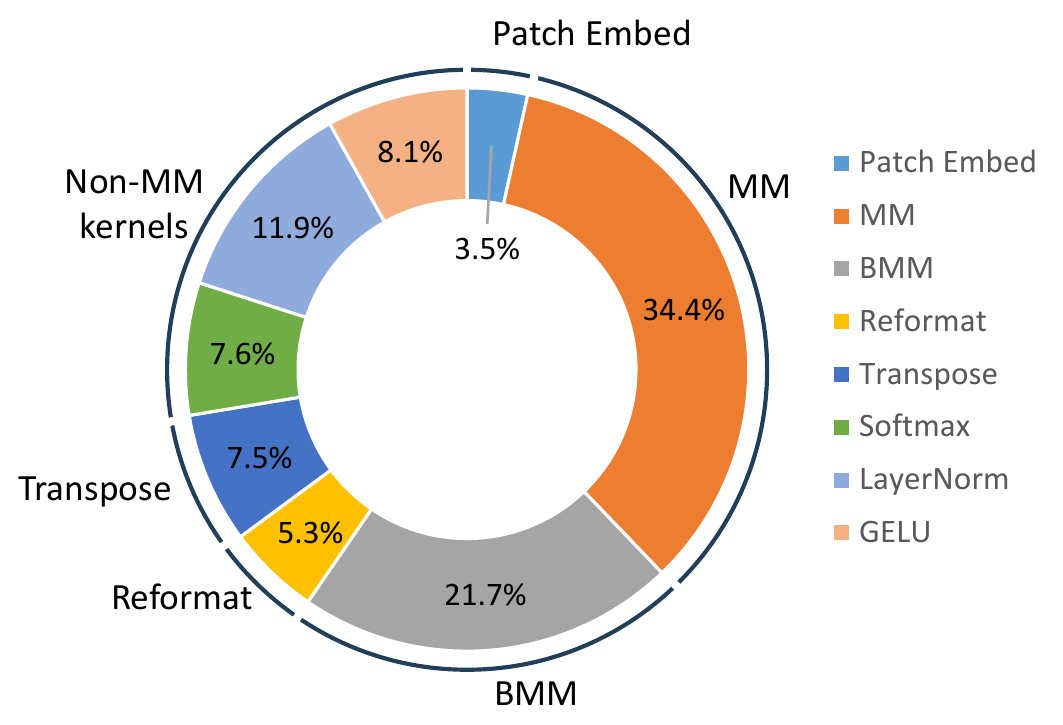}
\vspace{-15pt}
\caption{Kernel breakdown of DeiT-T inference latency on GPU A10G, batch size = 6. }
\label{fig:deit-t_profile_pie}
\end{figure}

Exploring latency throughput tradeoff requires a deep understanding of the performance.
To understand the performance of different layers within a deep learning application, we first perform an in-depth kernel profiling by using TensorRT~\cite{vanholder2016efficientTensorRT} to deploy an INT8 quantized DeiT-T inference on Nvidia GPU A10G. 
Built with 8nm fabrication, the Nvidia A10G GPU has 72 stream multiprocessor (SM)s with 4 tensor cores per SM, reaching the peak INT8 performance as 140 TOPS and peak FP32 performance as 35 TFLOPS, as specified in Table~\ref{tbl:a10_vck190_compare}.
We profile DeiT-T and set the batch size as 6.  
The measured end-to-end latency is 1.43 ms. 
We show the kernel time breakdown in Figure~\ref{fig:deit-t_profile_pie}.
We have the following observations:
\textbf{\raisebox{.5pt}{\textcircled{\raisebox{-.9pt} {1}}} 
The matrix-multiply or convolution-type kernel utilization is low.} 
This includes matrix-multiply (MM), batch matrix-multiply (BMM), and patch embedding, i.e., convolution.
We calculate the effective throughput in these layers as 
18 TOPS, which is only 13\% of
the peak INT8 throughput on A10G (140TOPS).
\textbf{\raisebox{.5pt}{\textcircled{\raisebox{-.9pt} {2}}} 
The nonlinear layers including Softmax, GELU, and LayerNorm take significant GPU cycles.}
These layers consume less than 1\% of the total computation operations, however, 
take around 28\% of the total time.
These layers are mapped to CUDA cores on the GPU.
\textbf{\raisebox{.5pt}{\textcircled{\raisebox{-.9pt} {3}}} 
The data layout change kernel consumes non-negligible GPU cycles, 
around 8\% of the total latency. 
}
The data layout change kernel, i.e., Transpose, is introduced either implicitly as certain data layouts are favorable for GPU Tensor Cores computation, e.g., the least dimension of the tensor is aligned with 32, or explicitly as specified in the model.
\textbf{\raisebox{.5pt}{\textcircled{\raisebox{-.9pt} {4}}}
The data type conversion kernel Reformat to convert between INT8 and FP32 also consumes non-negligible GPU cycles, around 5\% of the total latency. 
}
This happens, e.g., when the FP32 output from Softmax needs to be used as the input of the next matrix-multiply layer.

\input{tables/a10_vck190_compare}

We deploy the same INT8 quantized model, DeiT-T, on the AMD ACAP architecture~\cite{Versal_ACAP} VCK190~\cite{versal_vck190} board using CHARM~\cite{charmFPGA23}. 
CHARM~\cite{charmFPGA23} is the state-of-the-art deep learning inference accelerator and mapping framework on ACAP architecture, which features FPGA, AIE vector processors, and CPU on the system-on-chip. 
The end-to-end latency when using CHARM~\cite{charmFPGA23} is 12ms, 8.4x larger than that of GPU A10G under batch size 6. 
The main reason is that CHARM maps heterogeneous accelerators on ACAP and the data transfer among accelerators has to go to/from off-chip DDR. 
As specified in Table~\ref{tbl:a10_vck190_compare}, the VCK190 board has 25.6 GB/s off-chip bandwidth, which is much smaller than that of A10G. 
\textbf{\raisebox{.5pt}{\textcircled{\raisebox{-.9pt} {5}}}
Programming on ACAP creates new unsolved challenges.} 
Without careful design, performance on the ACAP will be constrained by the off-chip communication among accelerators, which leads to longer latency.

\begin{figure}[h]
\vspace{-5pt}
\centering
\includegraphics[width=1\columnwidth]{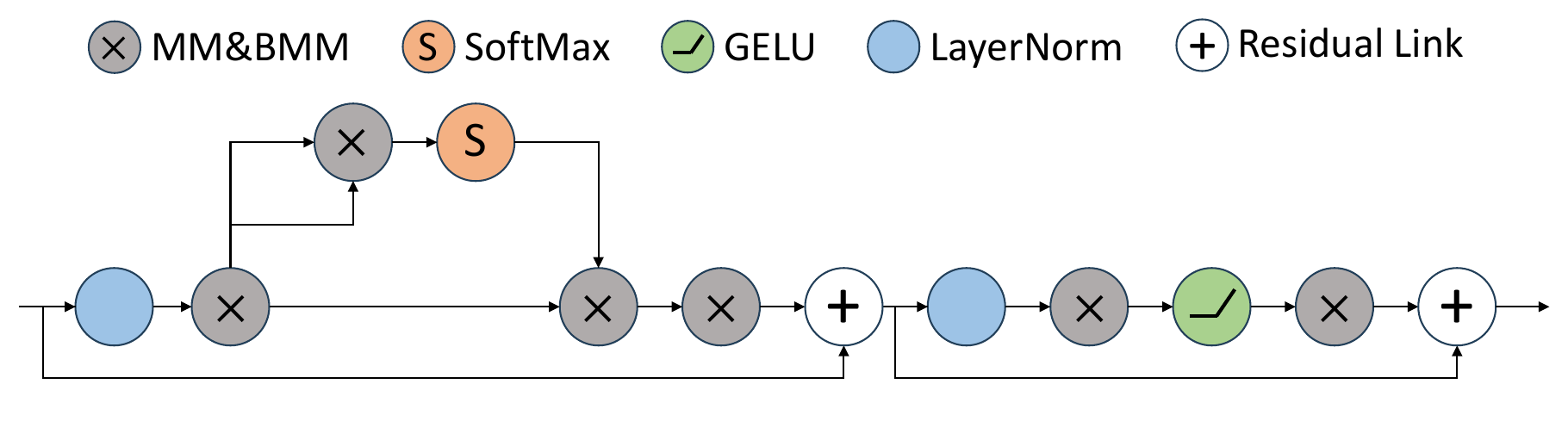}
\vspace{-25pt}
\caption{Layers \& their dependencies in a transformer block.}
\label{fig:transformer_block_dependency}
\vspace{-10pt}
\end{figure}

We further plot the layers within a transformer block in DeiT-T and show the dependencies between different layers in Figure~\ref{fig:transformer_block_dependency}. 
When considering the sequential spatial hybrid strategies, we can consider mapping different layers on one physical accelerator. 
For example, we can map all MM and batch MM layers using one MM accelerator and map all the other non-MM layers to separate accelerators. 
Using only one MM accelerator can potentially give us the lowest achievable latency for MM layers as discussed in Section~\ref{sec:introduction}. 
However, we should also consider the data communication between this one MM accelerator and all the other non-MM accelerators.
For example, the dataflow design and input \& output data layout design of this MM accelerator should be carefully chosen. 
Otherwise, it could be the case that the data layout of this MM accelerator matches with one neighboring non-MM accelerator but it does not match another one.
Therefore, it needs data layout change, which means extra communication overhead in addition to the computation of each layer.
Therefore, \textbf{\raisebox{.5pt}{\textcircled{\raisebox{-.9pt} {6}}}
when considering sequential spatial hybrid strategies, the data dependencies in the graph will make the communication patterns between accelerators more complex, and the inter-acc communication should be co-optimized during the accelerator design time.} 

To tackle these challenges, we propose ~\M~ to optimize performance, which brings the latency of mapping DeiT-T on VCK190 from 12 ms to 0.54 ms when batch size is 6, achieving a 22.22x speedup. 
Our ~\M~ solution beats the latency of GPU A10G by 2.53x. 
~\M~ also enables efficient latency throughput tradeoff design space exploration as described in Section~\ref{sec:introduction}.
How does ~\M~ achieve this?
\textbf{First}, ~\M~ explores sequential spatial hybrid strategies when mapping MM and BMM layers to enable the latency and throughput tradeoff.
~\M~ map these layers onto the AIE part of ACAP.
\textbf{Second}, ~\M~ considers on-chip forwarding when the model size fits on-chip. 
This greatly reduces the communication. But it also means the design complexity of the on-chip buffers increases. We discuss how to apply ~\M~ in general cases when the model size does not fit on-chip in Section~\ref{sec:insight}.
\textbf{Third}, ~\M~ designs efficient accelerators for nonlinear layers (Softmax, GELU, and LayerNorm), data layout change (Transpose), and data type conversions (Reformat) on the FPGA part. The flexibility provided by FPGA enables customization for various types of non-MM layers, which GPU CUDA cores lack.
\textbf{Fourth}, ~\M~ enables a fine-grained pipeline between MM layers on the AIE and non-MM layers on the FPGA to hide the non-MM latency, which further reduces the latency.
\textbf{Fifth}, ~\M~ considers the inter-acc communication during the layer-to-accelerator mapping stage and also the accelerator design stage. 
This further reduces the inter-acc communication overhead.

\input{tables/prior_compare}

%% file: tables/a10_vck190_compare.tex

\begin{table}[h]
\vspace{-5pt}
\caption{Comparisons between Nvidia GPU A10G and AMD Versal ACAP VCK190 on peak FP32 and INT8 performance, and peak off-chip bandwidth (BW).}
\vspace{-10pt}
\label{tbl:a10_vck190_compare}
\begin{adjustbox}{width=1\columnwidth,center}
\begin{tabular}{c|c c c }
\hline
Hardware Specification & FP32  & INT8 & Off-chip BW \\
\hline
Nvidia GPU A10G~\cite{a10gpu} &  35 T & 140 T & 600 GB/s \\
AMD ACAP VCK190~\cite{versal_vck190} & 6.4 T & 102.4 T & 25.6 GB/s \\
\hline
\end{tabular}
\end{adjustbox}
\end{table}

%% file: tables/prior_compare.tex
\begin{table*}
    \centering
\vspace{-10pt}
\caption{\textcolor{black}{Comparisons between SSR (ours) and prior works.}}
\vspace{-10pt}
\footnotesize
\begin{adjustbox}{width=2\columnwidth,center}
\begin{tabular}{c|c|cccccc}
\toprule
\multirow{2}{*}{\makecell[cc]{\textbf{Prior Works}}}  & \multicolumn{1}{c|}{\makecell[c]{\textbf{Computing} \textbf{Platform}} }    &\multicolumn{6}{c}{\textbf{Architecture Features}}  \\ 
\cline{2-8}
  &\textbf{Type}   & \makecell[cc]{\textbf{Spatial}\\\textbf{Accelerator}} &  \makecell[cc]{\textbf{Hardware}\\\textbf{Specialization}} &  \makecell[cc]{\textbf{On-chip}\\\textbf{Forwarding}}  & \makecell[cc]{\textbf{Fine-grained }\\\textbf{Pipeline}}& \makecell[cc]{\textbf{Hybrid}} & \makecell[cc]{\textbf{Inter-acc Comm.}\\\textbf{\&Acc Co-Design}} \\
\midrule

\makecell[cc]{{TensorRT}~\cite{vanholder2016efficientTensorRT}} &\textcolor{black}{GPU}  &\textcolor{red}{$\times$} &\textcolor{red}{$\times$} &\textcolor{red}{$\times$}&\textcolor{red}{$\times$} &\textcolor{red}{$\times$} &\textcolor{red}{$\times$}  \\
\arrayrulecolor{black!30}\midrule

\makecell[cc]{{DiviML}~\cite{ghannane2023diviml}} &{CPU+GPUs} &\textcolor{ForestGreen}{\checkmark}&\textcolor{ForestGreen}{\checkmark}&\textcolor{red}{$\times$}&\textcolor{red}{$\times$}&\textcolor{ForestGreen}{\checkmark} &\textcolor{red}{$\times$}   \\
\midrule


\makecell[cc]{{MAGMA}~\cite{kao2022magma}, {Herald}~\cite{kwon2021heterogeneous}} &ASIC &\textcolor{ForestGreen}{\checkmark} & \textcolor{ForestGreen}{\checkmark} & \textcolor{red}{$\times$} &\textcolor{red}{$\times$}&\textcolor{ForestGreen}{\checkmark}&\textcolor{red}{$\times$}     \\
\midrule


\makecell[cc]{{ViTCoD}~\cite{you2023vitcod}} &{ASIC} &\textcolor{red}{$\times$}&\textcolor{ForestGreen}{\checkmark}&\textcolor{red}{$\times$}&\textcolor{red}{$\times$}&\textcolor{red}{$\times$} &\textcolor{red}{$\times$}   \\
\midrule

\makecell[cc]{{SET}~\cite{MaKaishengISCA23}} &{ASIC}&\textcolor{ForestGreen}{\checkmark}&\textcolor{ForestGreen}{\checkmark}&\textcolor{ForestGreen}{\checkmark}&\textcolor{red}{$\times$} &\textcolor{ForestGreen}{\checkmark}&\textcolor{red}{$\times$}    \\
\midrule

\makecell[cc]{{HeatViT}~\cite{dong2023heatvit}, Auto-ViT-Acc}~\cite{lit2022auto} &{FPGA } &\textcolor{red}{$\times$}&\textcolor{ForestGreen}{\checkmark} &\textcolor{red}{$\times$}&\textcolor{red}{$\times$}&\textcolor{red}{$\times$} &\textcolor{red}{$\times$}  \\
\midrule

\makecell[cc]{{BrainWave}~\cite{putnam2014reconfigurable,caulfield2016cloud,firestone2018azure,fowers2018configurable,boutros2020beyond}, Intel NPU~\cite{boutros2020beyond}} &{FPGA }&\textcolor{ForestGreen}{\checkmark}&\textcolor{ForestGreen}{\checkmark}&\textcolor{ForestGreen}{\checkmark}&\textcolor{red}{$\times$}&\textcolor{red}{$\times$} &\textcolor{red}{$\times$}  \\
\midrule



\makecell[cc]{{DNNExplorer}~\cite{zhang2020dnnexplorer}} &{FPGA 
} &\textcolor{ForestGreen}{\checkmark}&\textcolor{ForestGreen}{\checkmark}&\textcolor{ForestGreen}{\checkmark}&\textcolor{red}{$\times$}&\textcolor{ForestGreen}{\checkmark} &\textcolor{red}{$\times$}   \\
\midrule

\makecell[cc]{{CHARM}~\cite{charmFPGA23}} &{ACAP } & \textcolor{ForestGreen}{\checkmark} & \textcolor{ForestGreen}{\checkmark}  &\textcolor{red}{$\times$} &\textcolor{ForestGreen}{\checkmark} &\textcolor{ForestGreen}{\checkmark}&\textcolor{red}{$\times$}   \\


\arrayrulecolor{black}\specialrule{0.75pt}{1pt}{3pt}

\makecell[cc]{\textbf{{SSR}}\\\textbf{(Ours)}} &{ACAP and FPGA}  &\textcolor{ForestGreen}{\checkmark}&\textcolor{ForestGreen}{\checkmark}&\textcolor{ForestGreen}{\checkmark}&\textcolor{ForestGreen}{\checkmark}&\textcolor{ForestGreen}{\checkmark}&\textcolor{ForestGreen}{\checkmark} \\
\arrayrulecolor{black}\specialrule{0.75pt}{1pt}{3pt}

\end{tabular}
\end{adjustbox}
\label{tbl:Prior_Comparison}
\vspace{-10pt}
\end{table*}

%% file: 3_RelatedWork.tex
\vspace{-10pt}
\section{Related Work}
\label{sec:relatedWork}
In this section, we first introduce existing approaches of sequential, spatial, and hybrid accelerators in Sections~\ref{subsec:Sequential Accelerator}, \ref{subsec:Spatial Accelerators}, \ref{subsec:Hybrid Accelerators}, and discuss their key features. 
We then summarize the comparisons between ~\M~ and the prior works in Table~\ref{tbl:Prior_Comparison}.
\vspace{-10pt}
\subsection{Sequential Accelerators}
\label{subsec:Sequential Accelerator}

GPUs are typically used as sequential accelerators in frameworks such as Tensorflow~\cite{TensorFlow}, Pytorch~\cite{PyTorch}, etc. With a lot of computing resources, GPUs achieve high throughput by batch processing. TensorRT~\cite{vanholder2016efficientTensorRT} provides general solutions for mapping deep learning models on GPUs. 
However, it does not provide customization on certain model workloads. Gemmini~\cite{gemmini-dac} is an automatic accelerator generator. It can generate both systolic-array-based and parallel vector engines like hardware accelerators. Gemmini has been widely applied to deep learning acceleration. 
For example, Sehoon et. al. ~\cite{kim2023full} use Gemmini in Transformer inference. The authors identify the characteristics of Transformer-based models and propose various optimization methods.
ViTCoD~\cite{you2023vitcod} designs a dedicated accelerator for sparse and dense workloads to boost hardware utilization for vision transformers. 
Auto-ViT-Acc~\cite{lit2022auto} designs an FPGA accelerator for multi-head attention and an FPGA-aware quantization algorithm to make better use of FPGA resources. 
HeatViT~\cite{dong2023heatvit} accelerates vision transformer on embedded FPGAs using image-adaptive token pruning and 8-bit quantization. 
However, these sequential accelerators use a generic accelerator for all layers with different shapes, which possibly leads to shape mismatch and results in larger latency. 
\vspace{-10pt}

\subsection{Spatial Accelerators}
\label{subsec:Spatial Accelerators}
Different from deep learning training, real-time AI inference applications usually do not have large batching inputs to fully explore parallelism, and therefore, many throughput-optimized systems for batch processing can only use a small portion of resources for a single inference request.
Microsoft BrainWave~\cite{putnam2014reconfigurable,caulfield2016cloud,firestone2018azure,fowers2018configurable} targets real-time AI inference in the data center scale production system. 
It explores parallelism within a single task and achieves much lower latency on FPGAs compared with GPUs without sacrificing system-level throughput.
Andrew et. al. ~\cite{boutros2020beyond} identify the gap between hardware's peak performance and achievable performance in real applications on Intel Stratix 10 NX FPGA. 
To minimize this gap in small batch AI inference, they re-implement BrainWave~\cite{putnam2014reconfigurable,caulfield2016cloud,firestone2018azure,fowers2018configurable} and propose enhanced neural processing unit (NPU) architecture on Intel Stratix 10 NX FPGA. 
By leveraging the flexibility of FPGA, they achieve significantly higher hardware utilization over GPUs with a comparable peak performance. 

\vspace{-10pt}
\subsection{Hybrid Accelerators}
\label{subsec:Hybrid Accelerators}
DNNExplorer~\cite{zhang2020dnnexplorer} proposes a hybrid design methodology. Specifically, applying spatial accelerators for the first several layers and using a generic accelerator for the rest layers to enable deep networks while achieving acceptable performance. DNNExplorer only supports a fine-grained pipeline between linear kernels, which can reduce latency to a certain extent, while in our work, we extend the pipeline to nonlinear kernels to further reduce end-to-end latency.
SET~\cite{MaKaishengISCA23} is a framework that automatically schedules deep neural network (DNN) nodes onto tiled accelerators. 
SET proposes a universal notation and formally defines the mapping space for analyzing tradeoffs among different schedule choices. However, it assumes a very flexible Network-on-Chip (NoC) to connect the accelerators which consumes non-negligible resources and may cause large overhead because of the data congestion in the NoC.
CHARM~\cite{charmFPGA23} composes heterogeneous accelerators for deep learning applications on ACAP. 
However, CHARM does not support on-chip data forwarding which results in longer inference latency.
DiviML~\cite{ghannane2023diviml} formalizes the DNN partition problem on the heterogeneous computing systems in which different accelerators such as GPUs are connected through PCIe links. 
DiviML proposes a linear programming model to search for both model and data parallelism and a heuristic schedule algorithm to optimize both latency and throughput. 
However, in DiviML, data transfer only happens after one layer finishes its computation, and overlap between computation and communication is not supported.
Herald~\cite{kwon2021heterogeneous} and MAGMA~\cite{kao2022magma} optimize DNN on heterogeneous computing systems, but different accelerators can only communicate with each other via off-chip memory, resulting in high latency.

We summarize the comparisons of ~\M~ with prior works in Table~\ref{tbl:Prior_Comparison}. ~\M~ 
adopts sequential spatial hybrid strategies, enables more scheduling flexibility to map layers to accelerators, designs fine-grained pipelines across different types of accelerators, and co-optimizes inter-acc communication with accelerator designs.
All together, ~\M~ achieves a better latency throughput Pareto front.

\vspace{-5pt}

%% file: 4_SSR_Methodology.tex
\begin{figure}[tb]
\centering
\includegraphics[width=1\columnwidth]{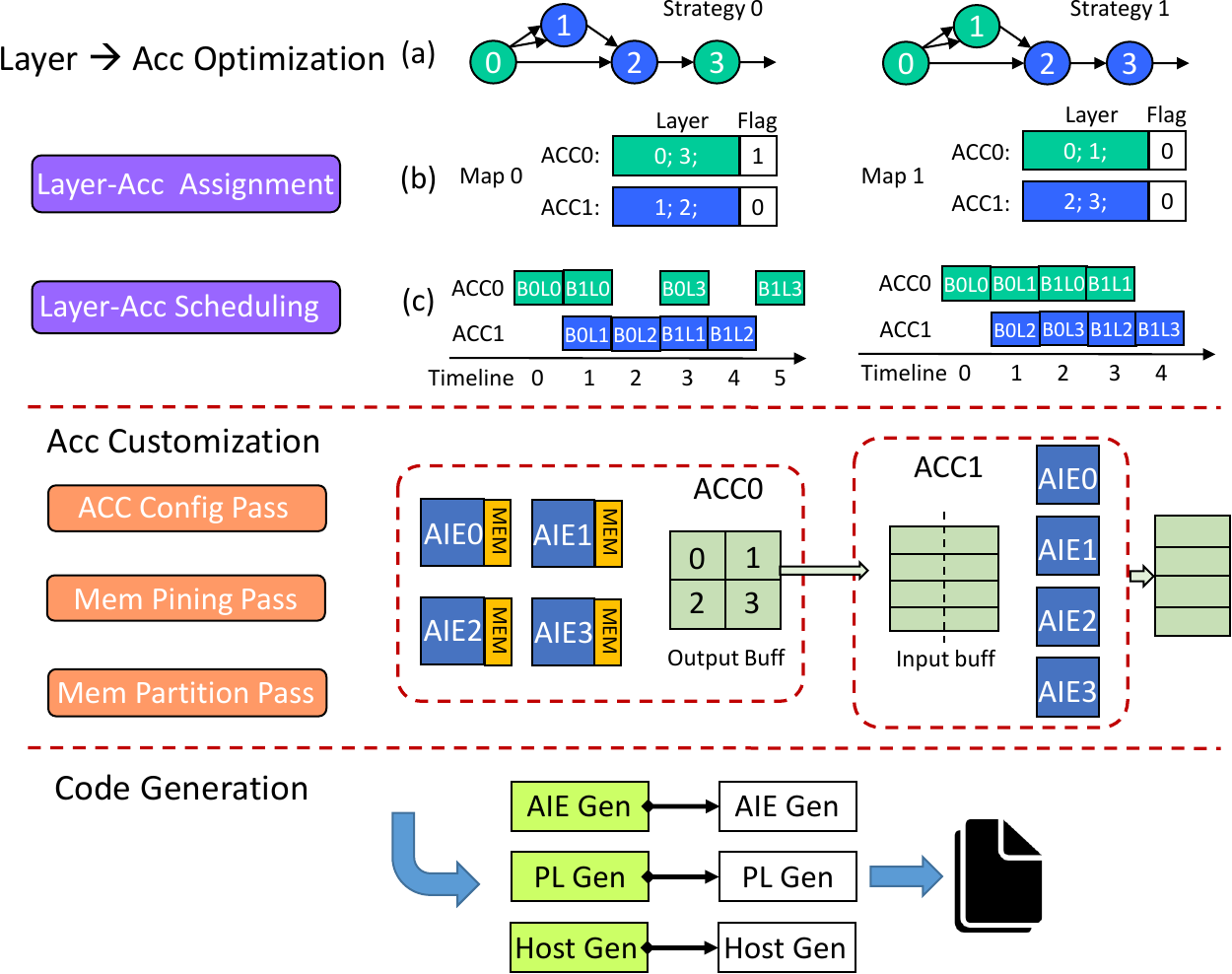}
\vspace{-20pt}
\caption{SSR framework overview.}
\label{fig:flow}
\end{figure}

\section{~\M~ Accelerator Architecture and ~\M~ Design Framework}
\label{sec:method}
In this section, we first introduce ~\M~ framework and heterogeneous architecture overview in \ref{sec:framework_overview} and \ref{sec:arch_overview}.
We then discuss hardware design methodologies and how to do efficient design space exploration in Sections~\ref{sec:design_method} and~\ref{sec:dse}.
Section~\ref{sec:acg} discusses code generation and compilation flow.
\subsection{SSR Framework Overview}
\label{sec:framework_overview}
Figure~\ref{fig:flow} illustrates the proposed ~\M~ framework.
The automatic framework takes the transformer model and hardware resource constraints as input and generates the spatial sequential hybrid execution scheduling as well as the corresponding hardware implementation on the Versal ACAP heterogeneous system. 
Our ~\M~ framework systematically optimizes the system throughput under certain latency constraints through two levels of optimization including Layer$\rightarrow$Acc level and Acc-Customization level.

At the Layer$\rightarrow$Acc level, given an application graph, the Layer$\rightarrow$ Acc scheduler will first generate the layer-accelerator assignment map by partitioning the graph into multiple sub-graphs and allocating each one to a specific accelerator. \jz{For example, as shown in Figure~\ref{fig:flow}(a), the graph consists of four layers. 
In strategy 0 (left), layers \{0, 3\} are assigned to Acc0, and layers \{1, 2\} are assigned to Acc1.} Based on the different layer-accelerator assignment maps, the scheduler can determine the execution order of the nodes with the dependency in the application graph being resolved. \jz{Assume there are two batches of input, denoted by B0 and B1, in strategy 0, it requires 6 units of time to finish two batches. In contrast, strategy 1 (right), requires 5 unit time. 
When considering the actual time in each unit, the Acc-Customization plays an important role, thus it leads to a coupled Layer$\rightarrow$Acc/Acc-Customization problem.}
After the Layer$\rightarrow$Acc assignment and scheduling, our framework will allocate the initial resource allocation constraints on each accelerator. 
\jz{Then the Acc-Customizer will optimize the configuration of each accelerator including the AIE array design, memory pinning strategy ({\large\ding{182}}), and non-linear kernel fine-grained pipeline design ({\large\ding{183}}).  
Most importantly, to reduce the data transfer overhead between different accelerators, we apply an inter-acc communication and accelerator co-design and introduce a customized memory partitioning strategy ({\large\ding{184}}).} 
Guided by the configuration provided by the ~\M~ scheduler, the automatic code generator will generate the source code for the host CPU, PL, and AIE respectively.

\begin{figure*}[tb]
\centering
\includegraphics[width=1.8\columnwidth]{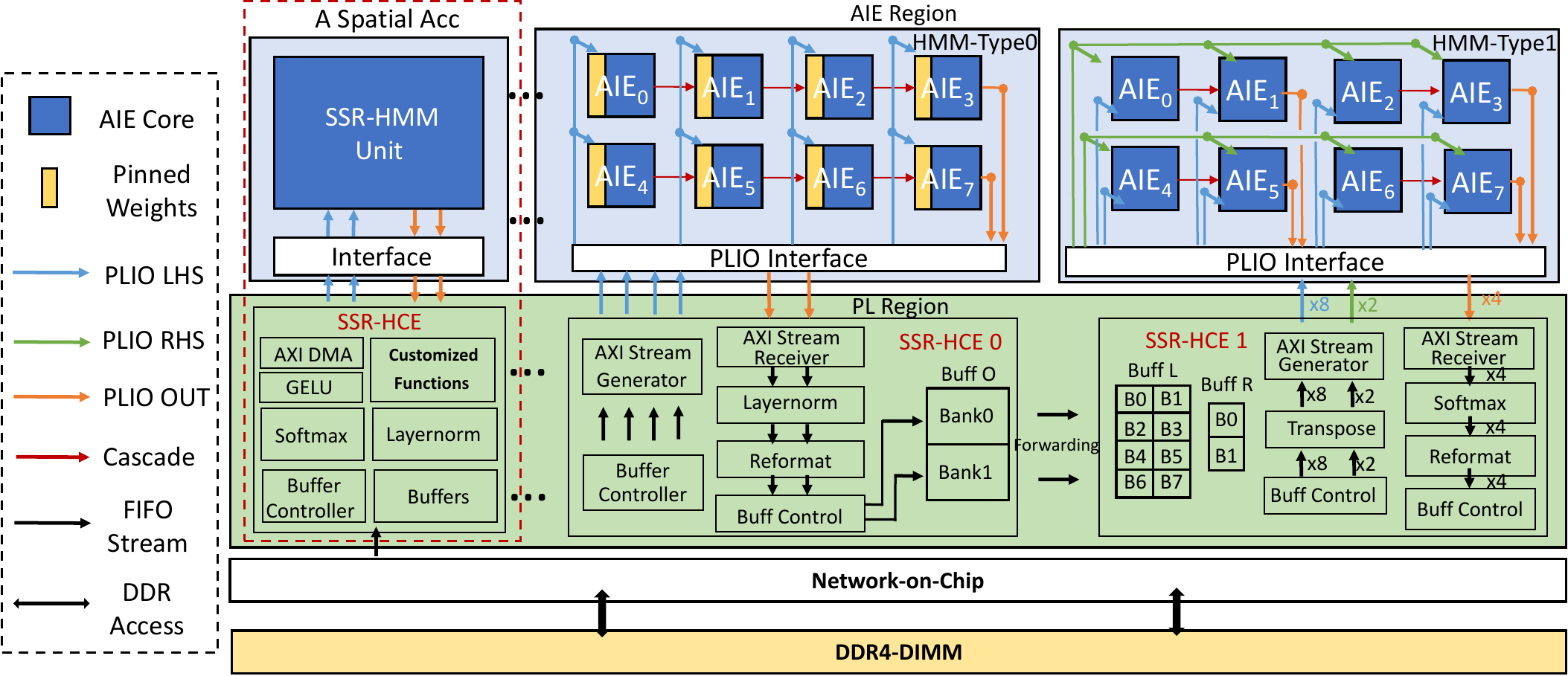}
\vspace{-10pt}
\caption{Proposed ~\M~ heterogeneous accelerator architecture overview.}
\label{fig:proposed_arch}
\vspace{-5pt}
\end{figure*}

\vspace{-10pt}
\subsection{~\M~ Heterogeneous Architecture Overview}
\label{sec:arch_overview}
\jz{The hardware architecture overview in our ~\M~ framework is shown in Figure~\ref{fig:proposed_arch}.}
It consists of N ($\in$ 1,...,n) spatial accelerators implemented on the AIE and PL. 
Within each spatial accelerator, there are two basic blocks, the heterogeneous matrix multiply (HMM) unit, and the heterogeneous customized engine (HCE). 

The AXI DMA in the spatial accelerator is responsible for sending the AXI request to the NoC that loads the image data/stores the final results from/to the off-chip DDR4 memory.
The HMM units handle the computation-intensive MM and BMM kernels using the high throughput AIE arrays. 
The HCE units contain senders and receivers to transfer the data between AIE and PL. 
The sender and receiver modules are not only responsible for generating the AXI stream protocols needed by the AIE array but also for computing the nonlinear and element-wise kernels. 
~\M~ supports extension for future applications as any customized function units can be included in our HCE units for data pre/post-processing.  
The intermediate data can move between different spatial accelerators through on-chip forwarding directly. 

\vspace{-10pt}
\subsection{SSR Hardware Design Methodology}
\label{sec:design_method}
\jz{After introducing the overall SSR architecture, we elaborate on the detailed hardware design methodology.
}\\
{\large\ding{182}}~\noindent\textbf{HMM configuration and memory pinning strategy.} 
\jz{In order to sustain the computation of 400 AIEs under the limited PLIO constraint~\cite{AutoMM}, we design two types of HMMs demonstrated in Figure~\ref{fig:proposed_arch}. 
For HMM-type0, by pinning the weights to the local memory of AIEs it only takes one operand (activations) to reduce the utilized PLIOs.} 
\jz{However, the multi-head attention layers in transformer models involve two activation operands, which cannot be implemented by HMM-type0. Thus HMM-type1 is designed to deal with such general matrix multiply operations.}
To apply the weights pinning and PLIO reduction strategy to the entire application graph, we mark each Layer$\rightarrow$Acc assignment with an optimizable flag. 
This is achieved by checking if attention layers are included in the assignment. 
For example, in Figure~\ref{fig:flow}(a), nodes 1 and 2 represent the multi-head attention layers with two activation inputs. 
When applying strategy 0, only non-attention layers are assigned to accelerator 0, thus we enable the optimization for searching the configuration to pin all the weights in the local memory of AIEs. 
By using this strategy, SSR enables high utilization of AIEs without routing congestion, for example, 394 AIEs out of a total of 400 AIEs are successfully implemented in the SSR-Spatial design.

\begin{figure}[tb]
\centering
\includegraphics[width=0.95\columnwidth]{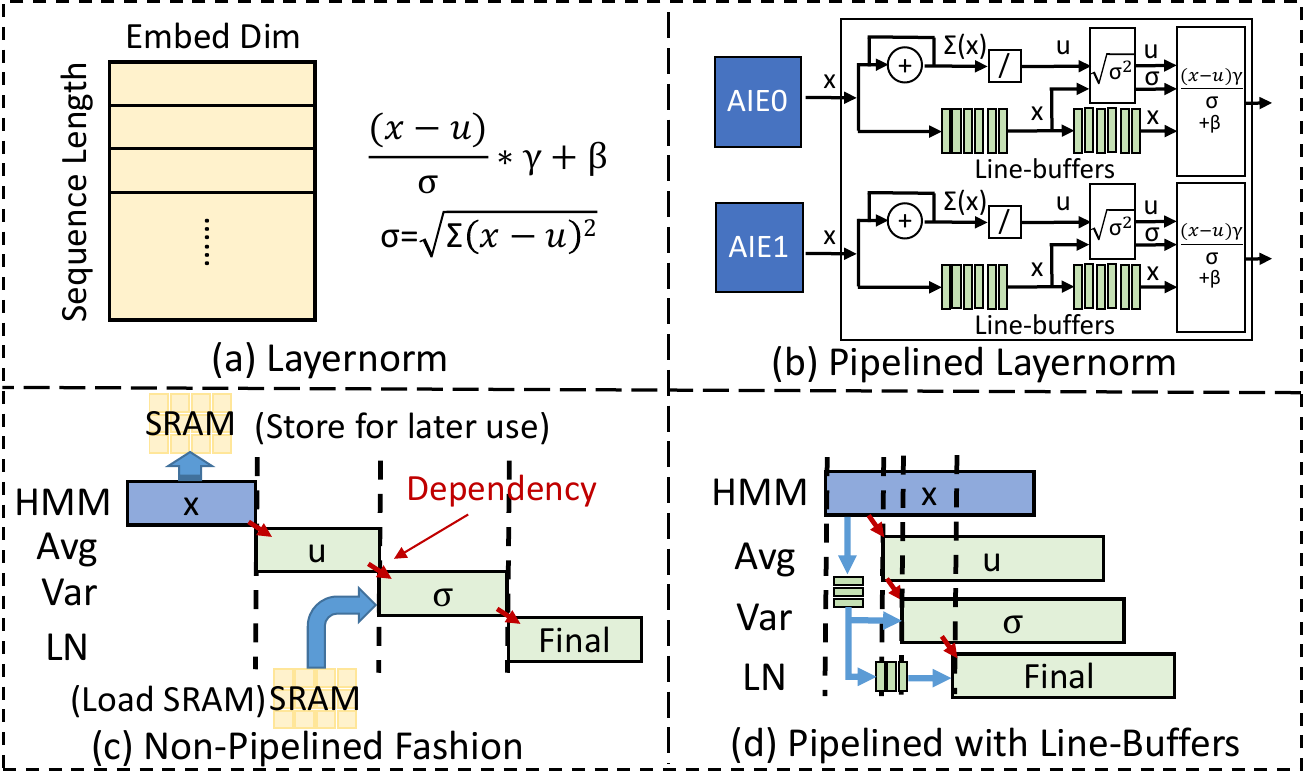}
\vspace{-10pt}
\caption{Element-wise and nonlinear kernel pipeline.}
\label{fig:kernel_pipeline}
\end{figure}

\noindent\textbf{\large\ding{183}~Fine-grained pipeline for element-wise and nonlinear kernels.} In order to reduce the latency of the non-computation-intensive kernels, we explore the fine-grained pipeline between the HMM and HCE units. 
The operations whose data reuse distance are one, such as Transpose, VectorAdd and
Reformat (data type conversion), can be easily fused with the HMM kernels. However, nonlinear operations such as Softmax, LayerNorm, and GeLU perform the reduction in an array resulting in the reuse distance larger than 1. 
Take the LayerNorm operation as an example as shown in  Figure~\ref{fig:kernel_pipeline}(a), before calculating the final results, it computes the average($\mu$) and standard derivation($\sigma$) along the embedding dimension. 
Moreover, the dependency also exists between average and standard derivation. If without any pipeline design, these operations can take even longer time compared to the computation-intensive HMM Units in Figure~\ref{fig:kernel_pipeline}(c). 
To reduce the latency and improve hardware utilization, we apply the bypass line-buffer structure in the customized Layernorm kernel on the PL side to overlap the latency in different stages as depicted in Figure~\ref{fig:kernel_pipeline}(b). 
As illustrated in Figure~\ref{fig:kernel_pipeline}(d), it receives data from HMM units and temporally pushes it into the line buffer. 
Right after the average $\mu$ of the first row is ready, it will read the data from the line buffer and calculate the standard derivation $\sigma$, so that the dependency can be resolved with a small waiting time. In general, this methodology can also be applied to other nonlinear kernels which reduces its latency to nearly half.

\begin{figure}[tb]
\centering
\includegraphics[width=1\columnwidth]{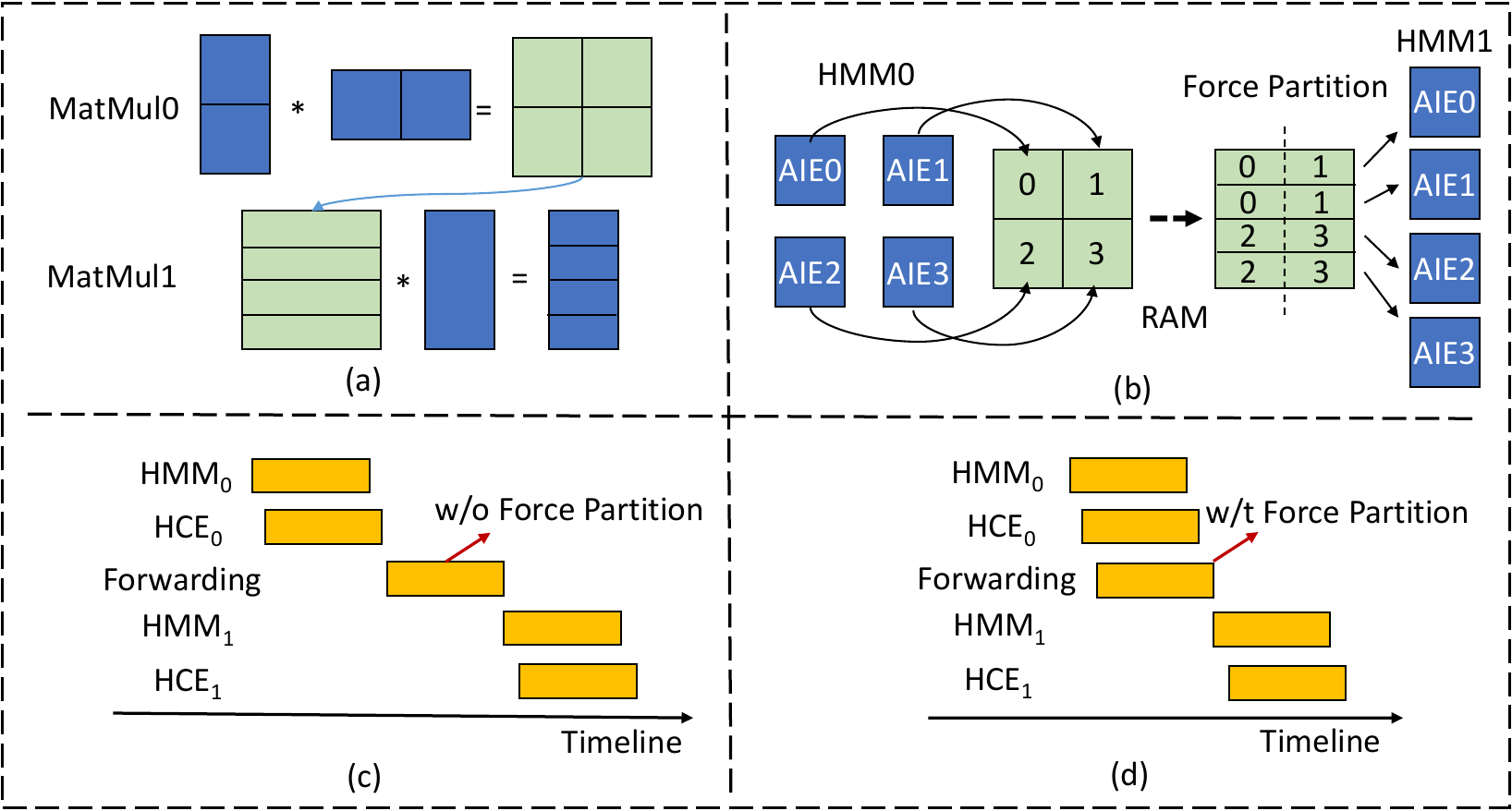}
\vspace{-20pt}
\caption{On-chip data forwarding between spatial accelerators with force RAM bank partition.}
\label{fig:force_partition}
\end{figure}

\noindent\textbf{\large\ding{184}~Inter-acc communication and accelerator co-design.}
When exploring the spatial-sequential architecture, the data communication patterns between accelerators become more complex and are prone to cause communication overhead because of the mismatch in accelerator configurations or memory conflicts. For example, the latency overhead appears in the consecutive matrix multiply scenarios as shown in Figure~\ref{fig:force_partition}(a) where MatMul0 and MatMul1 are mapped to HMM0 and HMM1 respectively. The output matrix of Matmul0 serves as the input activation of Matmul1.  
When designing the HMM kernels for MM with size M$\times$K$\times$N, there are three corresponding parallel choices at the AIE array level including here noted as A, B, and C. 
In other words, the A$\times$B$\times$C AIEs work concurrently with the A$\times$C AIEs generating the output at the same time. 
In Figure~\ref{fig:force_partition}(b), while \jz{HMM}0 parallels on A and C forming a 2$\times$2 AIE array, the \jz{HMM}1 parallels on A and B forming a 4$\times$1 AIE array. Since A$\times$C tiles of output will be transferred through PLIO and received by the downstream PL BRAM/URAM simultaneously, to prevent the HMM0 from stalling, A$\times$C bank partitioning is required shown in the RAM of HMM0. However, the data stored in the 2$\times$2 banks needs to be forwarded to the subsequent HMM1 as input activation in the format of 4$\times$1, resulting in bank conflicts. One straightforward solution to resolve bank conflicts is to introduce a non-overlapping operation that moves the data from RAM0 to RAM1 sequentially thus introducing a huge latency overhead in the pipeline as illustrated in Figure~\ref{fig:force_partition}(c). In this work, we propose a force-partition strategy to resolve the bank conflicts, while maintaining low latency. More specifically, during the runtime of optimization, we parse the data transaction among accelerators. For the pairs that do have communication, we configure the parallelism of the them to be divisible by each other. For example, the parallel parameters A, C of HMM0 should be fully divisible by the A, B of HMM1 or vice versa. Then we force the RAM bank partition of the subsequent \jz{HMM}1 to be compatible with the previous \jz{HMM}0. As illustrated in  Figure~\ref{fig:force_partition}(b), originally four banks of RAM are sufficient to guarantee the execution of the 4$\times$1 HMM1 unit. However only by partitioning the RAM to 4$\times$2, can the on-chip forwarding latency be overlapped by \jz{HMM}0 shown in Figure~\ref{fig:force_partition}(d).

\vspace{-10pt}

\input{Listing_DSE}
\subsection{~\M~ Design Space Exploration.}
\label{sec:dse}
\noindent\textbf{Layer$\rightarrow$Acc evolutionary algorithm (EA).} The main challenge to optimize the spatial-sequential solution is the extremely large design space. For example, the complexity for only Layer$\rightarrow$Acc scheduling is already over $O(9.9^n)$ ~\cite{MaKaishengISCA23} where $n$ is the number of layers in the graph. To solve this problem, we propose several heuristics at Layer$\rightarrow$Acc and Acc-Customization levels that explore the design space efficiently. At the Layer$\rightarrow$Acc level, we apply an evolutionary algorithm~\cite{holland1992genetic} based solution to optimize the throughput of the system while achieving the latency constraints demonstrated in Algorithm~\ref{code:inter_algo}. In our framework, the algorithm takes the execution graph, hardware resources, and latency constraints as input. By using the Layer$\rightarrow$Acc and Acc-Customization passes, it can generate the specialized configuration for each accelerator and the Layer$\rightarrow$Acc scheduling that will be used by our automatic generation to implement the design. The algorithm is inspired by the processes of biological evolution. It first randomly generates some Layer$\rightarrow$Acc strategies shown in Figure~\ref{fig:flow}(a-b) as the population and evaluates the design points in the current population through proposed SSR optimization passes ("SSR\_DSE" Lines 3-5). Then it selects the best assignment strategy to do crossover which generates the children generation (Lines 8-12). By introducing the mutation to the children generation it is possible to obtain a better assignment strategy (Lines 13-18). After evaluating all the design points, it will record the throughput optimal point under latency constraints and update the new population by selecting the top solutions (Lines 19-24). 

During the SSR Layer$\rightarrow$Acc and Acc-Customization processes (line 5 \& 18 defined in lines 27-37), by using a greedy algorithm, it first generates the Layer$\rightarrow$Acc scheduling pipeline and the data transaction between accelerators with the dependencies resolved according to the layer-accelerator mapping (Lines 28-29). More specifically, for a layer in the graph, we assign it to the pipeline as soon as its corresponding accelerator is available and its dependencies are already resolved as illustrated in Figure~\ref{fig:flow}(c). Then by analyzing the data transaction among accelerators, it determines a minimum memory allocation strategy that can buffer both the activations and weights on-chip while keeping the accelerator running without memory stall (Lines 30-31). Before doing the Acc-Customization (Lines 35-36, Algorithm~\ref{code:intra_algo}), the framework pre-allocates the resources to each accelerator including AIE, PLIO, RAM, and DSP. While the number of AIE together with PLIO is proportional to the total number of operations assigned to the accelerator, the memory budget is assigned according to the memory allocation strategy (Lines 32-33).

\input{Listing_intra}
\noindent\textbf{Inter-acc communication aware optimization at the Acc-Customization level.}
In the Acc-Customization stage, SSR searches the configurations of each accelerator represented as a config\_vector (h1, w1, w2, A, B, C, Part\_A, Part\_B, Part\_C). In the configuration, (h1, w1, w2) define the workload allocation per AIE, (A, B, C) determine the AIE array parallelism, and (Part\_A, Part\_B, Part\_C) determine the extra bank partitions for inter-acc communication aware optimization. In our design space, we find all integer solutions that make sure a single AIE workload can be fit in 32Kb AIE local memory and AIE utilization doesn't exceed the number of AIE. SSR sequentially launches the DSE for each accelerator according to the order of the accelerator appearing in the Layer$\rightarrow$Acc scheduling (Lines 2-4). This ensures that the other accelerators can get the information from the accelerators they depend on as much as possible. For example, for the first Layer$\rightarrow$Acc scheduling shown in Figure~\ref{fig:flow}, Acc0 will be searched before Acc1. Then SSR exhaustively searches the configuration of each accelerator within the design space defined before and makes sure the configurable meets the utilization constraints (Lines 6-11). The utilization can be calculated by Equation~\ref{eq:utilization} where the RAM\_util represents the number of RAMs needed in each partition and the DSP\_Util is the DSP utilization for each nonlinear processor. Then in order to avoid the communication overhead among accelerators due to the memory conflict problem discussed in Section~\ref{sec:design_method}, SSR takes two steps. First, it checks the AIE array configuration (A, B, C) of the current accelerator to align with the other accelerator that has data transactions. Then force memory bank partition is able to be launched (Line 12-18). The performance of each accelerator for its layers can be calculated by Equation~\ref{eq:perf}, since the nonlinear layers can be fully overlapped by MM kernels we omit it in the equation. After recording the configurable of each accelerator with the best performance (Line 20-23), it fine-tunes the communication overhead based on the knowledge of all the accelerators(Line 24).
\vspace{-15pt}

\begin{equation}
\begin{small}
\label{eq:utilization}
\begin{gathered}
    AIE = A*B*C \\
    PLIO = (A+C)*B \\
    RAM = Part_A*Part_B*Part_C*RAM\_util \\
    DSP = A*C*DSP\_util
\end{gathered}
\end{small}
\end{equation}

\begin{equation}
\begin{small}
\label{eq:perf}
\begin{gathered}
    Cycle = {{M*N*K}\over{A*B*C*MAC/Eff}}\\
    Throughput = {\#OPs\over{Cycle/Freq}}
\end{gathered}
\end{small}
\end{equation}

\subsection{Automatic Code Generation \& Compilation}
\label{sec:acg}
Our ~\M~ framework includes a Python interface to take model description as input and the output is the design source code files including ARM CPU host code, FPGA high-level synthesis code, and AIE intrinsic C/C++ code.
Based on our analytical model-guided design space exploration, the code generation toolflow can instantiate the code template to generate the design source code files.
~\M~ framework calls corresponding backend tools in AMD Vitis~\cite{vitis} 2021.1 to generate both the hardware bitstream and host binaries, which can be readily deployed on the board.

%% file: Listing_DSE.tex
\begin{algorithm}
\footnotesize
\begin{flushleft}
\textbf{Input}: Execution Graph (\textbf{G}), Hardware Constraints (\textbf{HW\_Cons}), Latency Constraints (\textbf{Lat\_Cons})\\

\textbf{Output}: SSR Spatial Acc Configuration (\textbf{Conf}), Layer-Acc scheduling (\textbf{schedule})

\textbf{Hyperparmeters}: \textbf{nAcc}, \textbf{nBat}, \textbf{nPop}, \textbf{nChild}, \textbf{nIter} \\
\Comment{\textcolor{magenta}{\textbf{nAcc} and \textbf{nBat} refers to the number of accelerators and batch of graphs, \textbf{nPop}, \textbf{nChild} and \textbf{nIter} are the parameters for EA search}} 
\end{flushleft}

\caption{SSR Evolutionary Algorithm}
\begin{minted}[fontsize=\footnotesize, linenos]{python}
assign_pop = zeros(nPop) #initialize layer-acc assignment
layer_acc_flag = 1 #enable inter-acc aware Acc-Customization
#Initialize first generation
assign_pop[:]=layer_acc_assign(nAcc)
latency, cost_thput_par[i], Conf, schedule=SSR_DSE(assign_pop[:],G)

for iter in range(nIter): #Run EA by nIter generations
    # Choose the best parent assignment and do single point crossover 
    for k in range(nChild//2):
        p1,p2 = assign_pop [select(cost_thput_par[:])]
        ch1,ch2 = sp_crossover(p1,p2)
        assign_chi.append(ch1,ch2)
    # Randomly exchange two layer-acc assignment to do mutation
    for k in range(nChild):
        assign_chi[k]=mutate(assign_chi[k])
        #Launch SSR optimization passes
        latency, cost_thput_chi[k], Conf, schedule = 
        SSR_DSE(assign_chi[k], G)
        if latency < Lat_Cons and cost_thput_chi[k]>best_thput:
            best_thput = cost_thput_chi[k]
            final_Conf, final_schedule = record(Conf, schedule)
    # Select top design points as new population 
    assign_pop = population_update (assign_pop, assign_chi)
    latency = cost_update (cost_thput_par, cost_thput_chi)
return final_Conf, final_schedule

def SSR_DSE (assign, Graph, layer_acc_flag):
    #Gready Algorithm based Layer->Acc scheduling
    acc_trans, schedule = layer_acc_schedule (assign, Graph)
    # First-round memory allocation based on data transfer among Accs
    mem_alloc = mem_allocation (acc_trans)
    # Determine AIE, PLIO, RAM, and DSP for each Acc
    hw_part = hw_partition (mem_alloc, schedule)
    # Launch SSR Acc-Customization DSE to get latency, throughput
    latency, thput, Conf = SSR_Acc_DSE (hw_part, schedule, ...
                          assign, acc_trans, layer_acc_flag)
    return latency, thput, Conf, schedule
\end{minted}
\label{code:inter_algo}
\end{algorithm}

%% file: Listing_intra.tex
\begin{algorithm}
\footnotesize
\begin{flushleft}
\Comment{\textcolor{magenta}{\textbf{hw\_part}, \textbf{schedule} and \textbf{acc\_trans} are described in Algorithm~\ref{code:inter_algo}. \textbf{hw\_part} contains the resource constraints for nAcc accelerators}} 
\caption{SSR inter-acc comm. aware customization}
\label{code:intra_algo}
\end{flushleft}
\begin{minted}[fontsize=\footnotesize, linenos]{python}
def SSR_Acc_DSE (hw_part,schedule,assign,acc_trans,inter_acc_flag)
    # Return the order for searching Accs
    index = trace_assignment(schedule)
    for i in index:
        final_thput = 0 #Initialize final throughput
        #exhaustive search the configuration in the design space
        for conf_vector[i] in Design_Space:
            util <-- Eq1 (conf_vector[i])
            #Check if resource utilization is under the constraints
            if util > hw_part[i]:
                continue
            #If inter-acc-aware is enabled, 
            if inter_acc_flag==1:
                #Check if the current configuration aligns with others
                if force_partition(conf_vector[i],assign)==false:
                    continue
                else: #Force memory partitioning to avoid overhead
                    update(conf_vector[i])
            cycle, thput <-- Eq2 (conf_vector[i],assign)
            if thput > final_thput:
                final_thput = thput
                final_cycle = cycle
                final_conf_vector[i] = conf_vector[i]
    final_cycle, final_thput= comm_overhead(final_cycle, schedule)
return final_cycle, final_thput
\end{minted}
\end{algorithm}

%% file: 6_SSR_Experiments.tex
\vspace{-5pt}
\section{Experiments}
\label{sec:results}
\subsection{Experimental Setup}

\input{tables/benchmarks}
We evaluate ~\M~ on AMD ACAP VCK190~\cite{versal_vck190} board with PL and AIE running on 230MHz and 1GHz respectively. We compare SSR with other state-of-the-art implementations of FPGA and GPU on four transformer-based applications shown in Table~\ref{tbl:model_setting}. The experiments setup for GPU, FPGA, and ACAP is summarized in
Table~\ref{tbl:exp_setup}. 
On GPU, we use ONNX 1.14.0 and TensorRT 6.1\cite{vanholder2016efficientTensorRT} to convert deep learning models from Pytorch and deploy inference with TensorRT. 
Then we measure the performance on Nvidia A10G GPU ~\cite{a10gpu} and use nvidia-smi~\cite{nvidia-smi} to measure the power consumption. 
On FPGA, we apply HeatViT~\cite{dong2023heatvit} on AMD Zynq ZCU102~\cite{ZCU102} and AMD Alveo U250~\cite{u250} as our baseline. 
AMD Board Evaluation and Management ~\cite{BEAM} is used to measure the power of 
ACAP boards.

\input{tables/exp_setup}

\input{tables/performance_energy_comparison}
\input{tables/latency_throughput_tradeoff}
\input{tables/Analytical_Model}
\vspace{-5pt}
\subsection{Performance \& Energy Efficiency Comparisons}

\subsubsection{Comparison of performance and energy efficiency among GPU, FPGA, and ACAP}
We apply the proposed SSR framework to four applications under three different batches. 
We verify the SSR designs on the AMD Versal VCK190 board and compare the latency, throughput, and energy efficiency with TensorRT~\cite{vanholder2016efficientTensorRT} solution on Nvidia A10G GPU, HeatViT~\cite{dong2023heatvit} solution on AMD ZCU102~\cite{ZCU102} and U250 FPGAs~\cite{u250}.

As shown in Table~\ref{tbl:performance and energy comparison}, ~\M~ outperforms all three other solutions under 3 different batch sizes in terms of latency, throughput, and energy efficiency. 
For SSR, the reported latency is measured when the number of accelerator(s) is set as the batch number. 
For all four applications with 3 different batch sizes of each, the average throughput gains ~\M~ achieves are 2.53x, 35.71x, and 14.20x when compared to Nvidia A10G GPU, AMD ZCU102, and U250 FPGA.
The average energy efficiency gains are 8.51x, 6.75x, and 21.22x, respectively. 
Specifically, when batch size = 1, the throughput gains are 2.84x, 21.67x and 9.38x, and the energy efficiency gains are 8.38x, 4.76x and 16.52x; 
when batch size = 3, the throughput gains are 2.37x, 35.54x and 14.05x, and the energy efficiency gains are 8.64x, 7.01x and 21.80x; 
when the batch size comes to 6, the throughput gains are 2.38x, 49.92x, and 19.18x, and the energy efficiency gains are 8.51x, 8.50x, and 25.35x, when compared to Nvidia A10G GPU, AMD ZCU102, and U250 FPGA respectively.

\subsubsection{Latency throughput tradeoff.}
In Table~\ref{tbl:latency_throughput}, we demonstrate the latency throughput tradeoff by comparing the throughput of A10G GPU, SSR-sequential design, SSR-spatial design, and SSR-hybrid design under certain latency requirements.  
In general, all the platforms achieve higher throughput when the latency constraints become looser. 
As described in Section~\ref{sec:introduction}, the GPU designs can only explore the latency throughput tradeoff by changing the batch size. 
Thus for the real-time scenarios with stringent latency constraints, e.g., <2ms as illustrated in Table~\ref{tbl:latency_throughput}, the small workload can't sustain the computation of GPU, and this results in relatively lower throughput. 
Moreover, GPU is unable to meet more critical latency requirements, e.g., <0.5ms. 

Since the SSR-spatial design is specialized for each layer in the application, it can achieve high computation utilization when the pipeline is filled with a sufficient number of batches. 
However, due to the resource partitioning, it has to sacrifice latency. Therefore it cannot meet the most critical time budget (<0.4ms). 
While the SSR-sequential design is capable of meeting all the latency constraints, due to the lack of specialization, it leads to shape mismatches between layers and the accelerator. Therefore it can't achieve high throughput. 
Among the design points, by adopting all the hardware optimization techniques and covering large design space, our proposed SSR-hybrid design is able to meet all the latency requirements and achieves the highest throughput under each latency constraint.

\subsubsection{Analytical modeling VS. On-board implementations}
We compare the latency of the DeiT-T model between the reported results by the SSR analytical model and the real on-board measurements in Table~\ref{tbl:analytical_model}. The design points are verified under the number of batches=6 with different numbers of accelerators. The error rate in percentage refers to the difference between the estimated latency by the SSR analytical model and the real on-board implementation. On average, the SSR modeling achieves less than 5\% error rate indicating that it can predict the hardware behavior accurately.

\begin{figure}[tb]
\centering
\includegraphics[width=0.7\columnwidth]{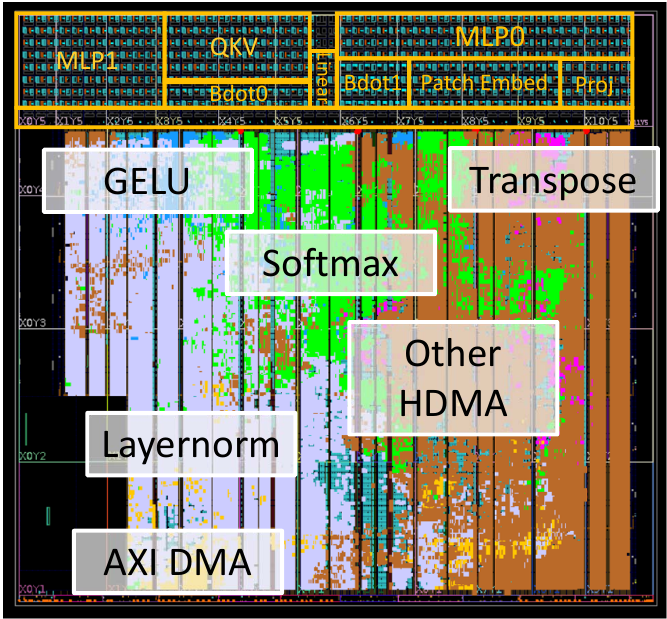}
\vspace{-5pt}
\caption{VCK190 implementation layout of ~\M-spatial. Each kernel is highlighted in the FPGA and AIE.}
\label{fig:implementation}
\vspace{-5pt}
\end{figure}

\input{tables/Hardware_Utilization}

\subsubsection{Implementation layout \& resource utilization breakdown}
The implementation layout of the proposed SSR-spatial design is shown in Figure~\ref{fig:implementation}. In this case, we design specialized MM accelerators on the AIE array for every node within one block of DeiT-T, e.g. QKV layer, attention layers, and MLP layers. The nonlinear kernels including layernorm, and softmax are implemented on the PL side. The corresponding hardware utilization breakdown is shown in Table~\ref{tbl:hardware_util} where specialized HMM units utilize 394 (98.5\%) AIEs and perfectly match the shape of layers in the DeiT-T model providing high AIE utilization. For the HCE units that support fine-grained pipeline, 799.8k (44.4\%) REG, 588.8k (65.4\%) LUT, 624 (64.5\%) BRAM, 104 (22.5\%) URAM and 1785 (90.7\%) DSPs are utilized.

\subsubsection{Search Efficiency}


\begin{figure}[tb]
\centering
\includegraphics[width=0.9\columnwidth]{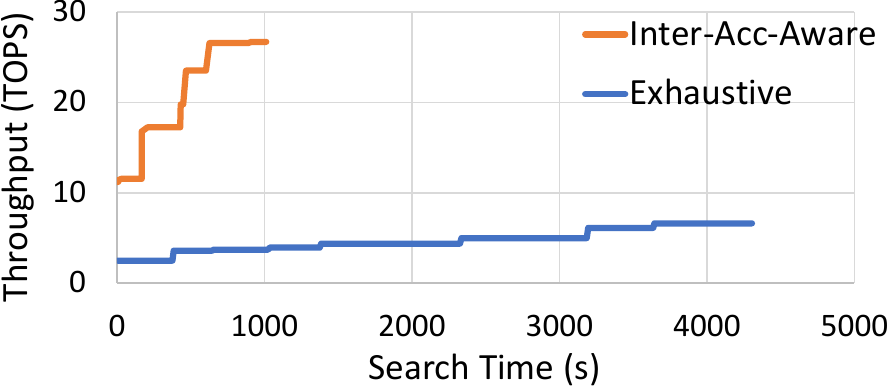}
\vspace{-5pt}
\caption{Search time comparison between inter-acc aware search and exhaustive search for DeiT-T.}
\label{fig:search_time}
\end{figure}

We apply the SSR design space exploration to optimize the throughput of the end-to-end inference under the latency constraints of less than 2ms. We compare the search efficiency of two proposed communication-aware strategies in Figure~\ref{fig:search_time}. The inter-acc aware strategy optimizes the communication overhead among accelerators by considering the configuration and bank partition of the other accelerators and thus is capable of pruning large inefficient design space. The baseline strategy exhaustively searches the design space and finally post-verifies the configuration of each accelerator and adds the communication overhead.
We conduct the search on an Intel Xeon Gold 6346 CPU utilizing 16 cores that run at 3.10GHz.
For DeiT, compared
to the baseline exhaustive search, the SSR inter-acc aware strategy finds the optimal solution of 26.70 TOPs within 1000s whereas the exhaustive search takes more than 4000s and still can not find high throughput designs.

\subsubsection{SSR Step-by-step optimization analysis}
SSR enables several design optimizations, including (1) on-chip data forwarding, (2) spatial accelerators, and (3) fine-grained pipeline. 
We measure the baseline design on VCK190 which none of the three optimizations is enabled. The latency of the baseline design is 12 ms for the DeiT-T model under batch=6, which is 22.2x slower than SSR 0.54 ms (ours). 
Compared to the baseline, when feature (1) is enabled, SSR achieves a 3.4x latency reduction on DeiT-T.  When feature (2) is enabled, it gives 2.4x more latency reduction. When feature (3) is further applied, SSR achieves another 2.7x latency reduction.

\vspace{-5pt}



%% file: tables/benchmarks.tex
\begin{table}[!tb]
\caption{Different vision transformer models configurations.}
\vspace{-10pt}
\label{tbl:model_setting}
\begin{adjustbox}{width=1\columnwidth,center}
\begin{tabular}{c|c c c c c}
\hline
\textbf{Model} & \textbf{\#Head} & \textbf{Embed. Dim} & \textbf{Depth} & \textbf{Model (M)} & \textbf{MACs (G)} \\
\hline
DeiT-T         & 3               & 192                 & 12             & 5.6                & 1.3               \\

DeiT-160       & 4               & 160                 & 12             & 4                  & 0.9               \\
DeiT-256       & 4               & 256                 & 12             & 7.4                & 2.1               \\
LV-ViT-T       & 4               & 240                 & 12             & 6.75                  & 1.6               \\
\hline
\end{tabular}
\end{adjustbox}
\vspace{-10pt}
\end{table}

%% file: tables/exp_setup.tex
\begin{table}
\begin{center}
\caption{Experimental hardware platforms.}
\vspace{-10pt}
\label{tbl:exp_setup}
\begin{adjustbox}{width=0.85\columnwidth,center}
    \begin{tabularx}{\linewidth}{ |c| *{2}{>{\centering\arraybackslash}X|}}

    \hline
    \multirow{5}{*}{GPU} & Board & NVIDIA A10G \\
    \cline{2-3}
    & Fabrication   & 8nm  \\
    \cline{2-3}
    & Frequency   & 1.71GHz  \\
    \cline{2-3}
    & TDP   & 300W  \\
    \cline{2-3}
    & Library   & TensorRT-8.6.1.6 \\
    \hline
    \multirow{5}{*}{FPGA} & Board & AMD U250\\
    \cline{2-3}
    & Fabrication   & 16nm  \\
    \cline{2-3}
    & Frequency   & 250MHz  \\
    \cline{2-3}
    & TDP   & 225W  \\
    \cline{2-3}
    \hline
    \multirow{5}{*}{FPGA} & Board & AMD ZCU102 \\
    \cline{2-3}
    & Fabrication   & 16nm  \\
    \cline{2-3}
    & Frequency   & 250MHz  \\
    \cline{2-3}
    & TDP   & 90W  \\
    \cline{2-3}
    \hline
    \multirow{5}{*}{ACAP} & Board & AMD VCK190 \\
    \cline{2-3}
    & Fabrication   & 7nm  \\
    \cline{2-3}
    & Frequency   & PL:230MHz, AIE:1GHz  \\
    \cline{2-3}
    & TDP   & 180W  \\
    \cline{2-3}
    \hline
    \end{tabularx}
\end{adjustbox}
\end{center}
\end{table}

%% file: tables/performance_energy_comparison.tex
\begin{table*}[]
\vspace{-5pt}
\caption{Performance and energy efficiency comparisons across different solutions.}
\vspace{-5pt}
\label{tbl:performance and energy comparison}
\begin{adjustbox}{width=1\textwidth,center}
\begin{tabular}{l|l|ccc|ccc|ccc|ccc}
\hline
                                   &                      & \multicolumn{3}{c|}{TensorRT~\cite{vanholder2016efficientTensorRT} on  A10G GPU} & \multicolumn{3}{c|}{HeatViT~\cite{dong2023heatvit}  on ZCU102} & \multicolumn{3}{c|}{HeatViT~\cite{dong2023heatvit}  on U250} & \multicolumn{3}{c}{~\M~ (ours) on VCK190} \\ \hline
model                              & Metrics              & Batch=1      & Batch=3     & Batch=6     & Batch=1    & Batch=3     & Batch=6    & Batch=1    & Batch=3   & Batch=6    & Batch=1    & Batch=3 & Batch=6          \\ \hline
\cellcolor[HTML]{FFFFFF}DeiT-T     & Latency (ms)           & 0.76         & 1.03        & 1.43        & 5.50        & 15.14      & 29.79      & 2.23       & 5.60        & 10.66       & 0.22      & 0.39      & 0.54      \\
                                   & Throughtput (TOPS)     & 3.19         & 7.05        & 10.16       & 0.44        & 0.48       & 0.49       & 1.09       & 1.30       & 1.36      & 10.90     & 18.62     & 26.70     \\
                                   & Energy   Eff (GOPS/W) & 26.54        & 40.76       & 48.37       & 46.82       & 48.96      & 49.25      & 14.02      & 16.66      & 17.04     & 246.15    & 368.75    & 453.32    \\ \hline
\cellcolor[HTML]{FFFFFF}DeiT-T-160 & Latency (ms)           & 0.73         & 1.05        & 1.45        & 4.22        & 11.81      & 23.18      & 2.21       & 5.67       & 10.88      & 0.21      & 0.37      & 0.50      \\
                                   & Throughtput (TOPS)     & 2.39         & 4.98        & 7.21        & 0.41        & 0.44       & 0.45       & 0.79       & 0.92       & 0.96      & 8.19      & 14.92     & 20.90     \\
                                   & Energy   Eff (GOPS/W) & 20.05        & 28.59       & 34.98       & 44.86       & 46.58      & 46.94      & 10.44      & 12.13      & 12.57     & 196.03    & 296.11    & 360.90    \\ \hline
\cellcolor[HTML]{FFFFFF}DeiT-T-256 & Latency (ms)           & 0.81         & 1.17        & 1.69        & 9.10        & 25.56      & 50.51       & 3.52       & 9.07       & 17.24     & 0.40      & 0.66      & 0.98      \\
                                   & Throughtput (TOPS)     & 5.09         & 10.56       & 14.63       & 0.45        & 0.48       & 0.49       & 1.17       & 1.36       & 1.43      & 10.30     & 18.73     & 25.22     \\
                                   & Energy   Eff (GOPS/W) & 38.53        & 51.78       & 66.78       & 543.55       & 46.48      & 46.16      & 15.05      & 17.43      & 18.27     & 229.37    & 363.59    & 423.89    \\ \hline
\cellcolor[HTML]{FFFFFF}LV-ViT-T   & Latency (ms)           & 0.92         & 1.37        & 1.91        & 7.24        & 20.27      & 39.95      & 3.11       & 7.91       & 15.11      & 0.38      & 0.62      & 0.85      \\
                                   & Throughtput (TOPS)     & 3.39         & 6.84        & 9.81        & 0.43        & 0.46       & 0.47       & 1.01       & 1.18       & 1.24      & 8.21      & 15.10     & 22.03     \\
                                   & Energy   Eff (GOPS/W) & 21.34        & 35.79       & 45.19       & 43.97      & 46.20      & 45.52      & 12.53      & 14.69      & 15.32     & 181.74    & 296.74    & 360.04    \\ \hline
\end{tabular}
\end{adjustbox}
\vspace{-5pt}
\end{table*}


%% file: tables/latency_throughput_tradeoff.tex
\begin{table}[!tb]
\caption{Comparisons on the optimal throughput (TOPS) under four different latency constraints (ms) for four solutions including TensorRT on GPU A10G, and SSR designs (ours) on VCK190 for DeiT-T. SSR-hybrid includes designs from SSR-sequential and SSR-spatial.}
\vspace{-5pt}
\begin{adjustbox}{width=0.9\columnwidth,center}
\begin{tabular}{*{5}{c}}
\toprule
\makecell[c]{Latency\\ Constraints} & \makecell[c]{GPU\\ (TensorRT)} & \makecell[c]{SSR-\\sequential\\ (ours)} & \makecell[c]{SSR-\\spatial\\ (ours)} & \makecell[c]{SSR-\\hybrid\\ (ours)} \\ \midrule
2 ms               & 11.32      & 11.17                & 26.70               & 26.70              \\
1 ms               & 5.28       & 11.12                & 26.70               & 26.70              \\
0.5 ms             & \textcolor{red}{$\times$}         & 11.05               & 19.37          & 19.37      \\
0.4 ms            & \textcolor{red}{$\times$}         & 10.90                  & \textcolor{red}{$\times$}                  & 18.56              \\ \bottomrule
\end{tabular}
\end{adjustbox}
\label{tbl:latency_throughput}
{\raggedright \small Note: \textcolor{red}{$\times$} means can not find a valid solution under the latency constraint.
}
\end{table}

%% file: tables/Analytical_Model.tex
\begin{table}[!tb]
\vspace{-5pt}
\caption{Latency comparison for DeiT-T between SSR analytical modeling and on-board measurements.}
\vspace{-10pt}
\label{tbl:analytical_model}
\begin{adjustbox}{width=0.9\columnwidth,center}
\begin{tabular}{c | c | c | c}
\toprule
\textbf{\textcolor{black}\# of Accs} & \textbf{\textcolor{black}Estimation(ms)} & \textbf{\textcolor{black}On-board(ms)} & \textbf{\textcolor{black}Error Rate} \\
\midrule
1  & 1.29 &1.30  &  1\%  \\
2  & 1.14 &1.08  & -6\%  \\
3  & 0.88 &0.85  & -4\%   \\
4  & 0.81 &0.83  &  3\%   \\
5  & 0.77 &0.79  &  2\%   \\
6  & 0.54 &0.54  & -1\%   \\
\bottomrule
\end{tabular}
\end{adjustbox}
\end{table}

%% file: tables/Hardware_Utilization.tex
\begin{table}[]
\caption{SSR hardware utilization for DeiT-T on INT8 mode.}
\vspace{-10pt}
\label{tbl:hardware_util}
\begin{adjustbox}{width=1\columnwidth,center}
\begin{tabular}{|c|c|c|c|c|c|c|c|}
\hline
\textbf{Modules}   &\textbf{REG}   & \textbf{LUT}   & \textbf{BRAM}   & \textbf{URAM} & \textbf{DSP}  & \textbf{PLIO}  & \textbf{AIE} \\ \hline
Total      &849527 & 619956 & 624 & 104 & 1797 & 199  & 394  \\ \hline
AXI DMA    &10316  & 5482   & 12  & 0   & 12   & --   & --  \\ \hline
Layernorm  &308736 & 256678 & 0   & 0   & 1024 & --   & --  \\ \hline
Softmax    &179544 & 78549  & 192 & 0   & 336  & --   & --  \\ \hline
GeLU       &3888   & 2400   & 0   & 0   & 0    & --   & --  \\ \hline
Transpose  &13541  & 5720   & 0   & 0   & 0    & --   & --  \\ \hline
Other HCE &333502 & 271127 & 420 & 104 & 425  & --   & --  \\ \hline
HMM        &0      & 0      & 0   & 0   & 0    & 199  & 394 \\ \hline
\end{tabular}
\end{adjustbox}
\end{table}

%% file: 7_Discussion.tex
\section{Discussion of Mapping Insights}
\label{sec:insight}


\noindent\textbf{Q1: Can we leverage ~\M~ in other architectures?} \\
\textcolor{black}{\noindent\textbf{A1: Yes. ~\M~ can be applied to other architectures.
}}\\
~\M~ can be used as a general solution and we can apply ~\M~ mapping method to other platforms, for example, Intel Stratix 10 NX FPGA~\cite{intel_fpga}, which has AI-optimized tensor blocks with up to 143 INT8 TOPS, 16MB on-chip memory, and 512GB/s high bandwidth memory. 
We use ~\M~ analytical models to estimate the latency after we change the hardware resource configurations to be fed into the modeling. 
In our modeling, we use data from \cite{boutros2020beyond} and \cite{fowers2018configurable} to get a reasonable INT8 computation efficiency for MM kernels and other non-MM kernels on Intel Stratix 10 NX. 
The modeled latency when adopting ~\M~ to map DeiT-T on Intel Stratix 10 NX FPGA is 0.49ms, which is comparable to 0.54ms on VCK190 (0.41m ms if VCK190 has 102GB/s off-chip bandwidth). 
This indicates one of the key contributions of ~\M~, i.e.,  ~\M~ provides a general mapping solution that can improve performance across platforms.

\noindent\textbf{Q2: Can we leverage ~\M~ when model sizes do not fit on-chip?} \\
\textcolor{black}{\noindent\textbf{A2: Yes. 
If a model can not fit on a single board, we can leverage ~\M~ to explore how the model is most effectively partitioned onto multiple devices.  
}}\\
Extensive works have discussed partitioning a large model onto multiple devices spatially whereas part of the model could fit onto the chip. 
Microsoft Catapult/Brainwave projects deploy large applications (machine learning, search engine, etc.) onto multiple directly connected FPGAs~\cite{putnam2014reconfigurable} within a server rack or onto a larger number of FPGAs connected with secondary rack-scale networks for inter-FPGA communication~\cite{caulfield2016cloud,firestone2018azure,fowers2018configurable}. 
Specifically, we can use a similar assumption as in~\cite{fowers2018configurable}, where the system stores deep learning models' weights in distributed on-chip SRAM memories. 
For example, the DeiT-Base model is 16x larger than DeiT-T in parameter size. 
According to the inter-FPGA latency reported in~\cite{fowers2018configurable,firestone2018azure}, we can scale out our design onto 12 VCK190 boards connected via 100Gb/s QSFP28 with 0.1 ms inter-FPGA board communication overhead across each board.

%% file: 8_Conclusion.tex
\section{Conclusion and Acknowledgement}
\label{sec:conclusion}
In this work, we propose ~\M~ accelerator \& ~\M~ framework to design the sequential spatial hybrid architecture to explore latency throughput tradeoff for deep learning applications and achieve a better Pareto front than sequential-only and spatial-only designs.

We acknowledge the support from NSF awards 2213701, 2217003, 2324864, 2328972, and the University of Pittsburgh New Faculty Start-up Grant.
We thank all the reviewers for their valuable feedback and AMD/Xilinx for hardware and software donations.